\documentclass[twocolumn,aps,pre,showpacs,preprintnumbers,floatfix,superscriptaddress]{revtex4-1}

\usepackage{amsmath}
\usepackage{amssymb}
\usepackage{graphicx}
			
\usepackage{xspace}
\newcommand{\eg}{{e.g.,\/}\xspace}
\newcommand{\ie}{{i.e.,\/}\xspace}
\newcommand{\etal}{{\it et~al.\/}\xspace}

\newcommand{\eq}[1]{(\ref{#1})}
\newcommand{\Eq}[1]{Eq.~(\ref{#1})}

\newcommand{\Fig}[1]{Fig.~\ref{#1}}
\newcommand{\Sec}[1]{Sec.~\ref{#1}}

\begin{document}

\title{Beyond nonlinear saturation of backward Raman amplifiers}

\author{Ido Barth}
\email{ibarth@princeton.edu} 
\affiliation{Princeton Plasma Physics Laboratory, Princeton University, Princeton, New Jersey 08543, USA}

\author{Zeev Toroker}
\affiliation{Department of Electrical Engineering, Technion Israel Institute of Technology, Haifa 32000, Israel}

\author{Alexey~A. Balakin}
\affiliation{Institute of Applied Physics RAS, Nizhnii Novgorod 603950, Russia}

\author{Nathaniel~J. Fisch}
\affiliation{Department of Astrophysical Sciences, Princeton University, Princeton, New Jersey 08540, USA}

\date{\today}

\begin{abstract}
Backward Raman amplification is limited by relativistic nonlinear dephasing resulting in saturation of the leading spike of the amplified pulse. 
Pump detuning is employed to mitigate the relativistic phase mismatch and to overcome the  associated saturation.
The amplified pulse can then be reshaped into a mono-spike pulse with little precursory power ahead of it,   with the maximum intensity increasing by a factor of two.
This detuning can be employed advantageously both in regimes where the group velocity dispersion is unimportant and where the dispersion is important but small.
\end{abstract}


\maketitle

\section{Introduction}
Laser intensities were significantly increased during the past decades, mainly by the means of chirped pulse amplification, up to the dielectric grating limitation \cite{CPA}.
The next generation of high intensity laser will require a different medium for amplification, such as plasma, which can tolerate much higher intensities and fluences. 
Particularly, backward Raman amplification (BRA) was suggested as such a scheme \cite{Malkin_PRL_99}, taking advantage of the resonant energy transfer between two counterpropagating lasers that interact via an electrostatic plasma wave.
The possibility of reaching nearly relativistic unfocused intensities in backward Raman amplifiers has been in principle demonstrated experimentally as well 
\cite{Ping_PRL_04,Balakin_JEPT_04,Cheng_PRL_05,Ren_PoP_08,Pai_PRL_08,Jaroszynski_PRL_05,Jaroszynski_NJP_11,Jaroszynski_SR_15}.

The major physical processes that may affect BRA include the amplified pulse filamentation 
\cite{Malkin_PRL_99,Malkin_EPJ_14,Fraiman_PoP_02,Malkin_PoP_12}, 
detuning due to the relativistic electron nonlinearity \cite{Malkin_PoP_14,Malkin_PRE_14}, 
parasitic Raman scattering of the pump and amplified pulses by plasma noise 
\cite{Malkin_PRL_99,Malkin_PoP_00,Malkin_PRL_00_a,Malkin_EPJ_14,Solodov_PRE_04}, 
generation of superluminous precursors of the amplified pulse 
\cite{Tsidulko_PRL_02}, 
pulse scattering by plasma density inhomogeneities 
\cite{Solodov_PoP_03}, 
pulse depletion and plasma heating through inverse bremsstrahlung \cite{Malkin_PRE_07,Balakin_PoP_11}, 
the resonant Langmuir wave Landau damping \cite{Malkin_PRE_07,Malkin_PoP_10,Hur_PRL_05,Depierreux_Nat_14} 
or breaking \cite{Malkin_PRL_99,Malkin_PoP_00,Malkin_EPJ_14,Trines_Nat_11,Toroker_PoP_14}, 
and other processes (see, for examples, 
Refs.~\cite{Clark_PoP_02,Yampolsky_PRE_04,Toroker_PoP_12,Toroker_PRL_12}). 
Most of these deleterious processes can be mitigated by proper preparation of laser pulses and plasmas, choosing parameter ranges and selective detuning of the Raman resonance. 
Ultimately, the output intensity limit appears to be imposed primarily by the relativistic electron nonlinearity, causing a phase mismatch of the Raman resonance that results in a saturation of the dominant leading spike growth \cite{Malkin_PoP_14} as well as of the secondary higher spikes in the multi-spike wave train solution \cite{Malkin_PRE_14}. 

It is the objective of the present study to address means of overcoming the saturation due to the phase mismatch caused by the relativistic nonlinearity.
To this end, we suggest to compensate the relativistic phase mismatch by proper detuning of the pump frequency.
This will delay the saturation and allow a longer resonant amplification of the leading spike, resulting in a reshaped, single-spike pulse of higher intensity.

Detuning of the pump frequency, accompanied by a density gradient, was suggested to suppress unwanted noise \cite{Malkin_PRL_00_a}, forward Raman scattering \cite{Malkin_PRL_00_b}, superluminous precursors \cite{Tsidulko_PRL_02}, and to get superradiant linear Raman amplification    \cite{Jaroszynski_PRL_05,Jaroszynski_NJP_11,Jaroszynski_SR_15}.
However, these chirps were employed only at early stages, before the pump encounters the seed or in the linear regime when the pump is not yet depleted.
In the advanced nonlinear stage, when the pumped signal is highly amplified to relativistic intensities, the effect of pump detuning remains to be studied. 
It is notable that all the aforementioned chirping advantages in the early stages can be included in the chirping suggested here, which mainly depends on the detuning profile at late times. 
The specific detuning, however, that will cure the relativistic nonlinearity has a more complicated functional dependency as derived here.
It will be found here that, by compensating for the phase mismatch through pump detuning, a factor of two in the maximal intensity can be achieved. 
Moreover, this intensity is attained with little leading power ahead of it.  

The plan of the paper is as follows:
In  Sec.~\ref{sec2}, we introduce the physical model, \ie 3-wave interaction equations including nonlinearity, dispersion, and detuning.
More details on the model are given in the Appendix.
In Sec.~\ref{sec3} we introduce a theory for the phase mismatch evolution in the dispersionless regime and compare with simulations.
Sec.~\ref{sec4} is devoted to the main results, the effect of detuning and a roadmap for chirp optimization.
In Sec.~\ref{sec5} we address the effect of dispersion on the detuned solution and suggest physical parameters for future experiments.
The deleterious influence of the secondary Raman scattering on the amplified pulse is estimated in Sec.~\ref{sec6}.
The conclusions are summarized in Sec.~\ref{sec7}.

\section{model} \label{sec2}
We adopt the dimensionless quasi-static 3-wave interaction model
including dispersion, detuning, and relativistic nonlinearity 
\cite{Malkin_PRL_00_a,Malkin_PRL_07,Malkin_PRE_14}  
(for detailed definitions and derivation, 
see Eqs.~(\ref{a_tilde})--(\ref{b_tilde})  in the Appendix),
\begin{eqnarray}
	a_{\zeta} &=& -b f,       \label{a_} \\
	f_{\zeta } &=& a b^{*} -i\Delta f,  \label{f_} \\
	b_{\tau } &=& a f^{*} - iQ \,b_{\zeta\zeta} +i|b|^2 \, b.  \label{b_} 
\end{eqnarray}
Here, $a$, $b$, and $f$ are proportional to the envelopes of the pump pulse, counterpropagating shorter seed pulse, and resonant Langmuir
wave, respectively; subscripts signify derivatives with respect to the elapsed amplification time, $\tau$, and the distance (or delay time), $\zeta$, from the original seed maximum, $\zeta_0$. 
The two parameters in the model are the rescaled dispersion coefficient, $Q$, and the rescaled detuning, $\Delta\sim\delta\omega=\omega_f+\omega_b-\omega_a$, where $\omega_{a,b,f}$ are the frequencies of the pump, seed, and Langmuir waves, respectively.
For strongly under-critical plasma, $\omega_e \ll \omega_a$, we have
$Q\approx \omega_e/2\omega_b$ (see \Eq{Q} in the Appendix).

Equations (\ref{a_})--(\ref{b_}) are solved numerically for initial constant pump $a(\zeta,\tau=0)=1$, zero Langmuir wave, $f(\zeta,\tau=0)=0$, and
small input Gaussian seed pulse of the form
\begin{equation}
	b(\zeta,\tau=0)=\frac{b_0}{\sqrt{D\pi}} \exp\left[-\frac{(\zeta-\zeta_0)^2}{D}\right] \label{b0}
\end{equation}
with $b_0=0.05$, $D=1$, and $\zeta_0=10$.

\begin{figure}[tr]
\includegraphics[width=9.0cm]{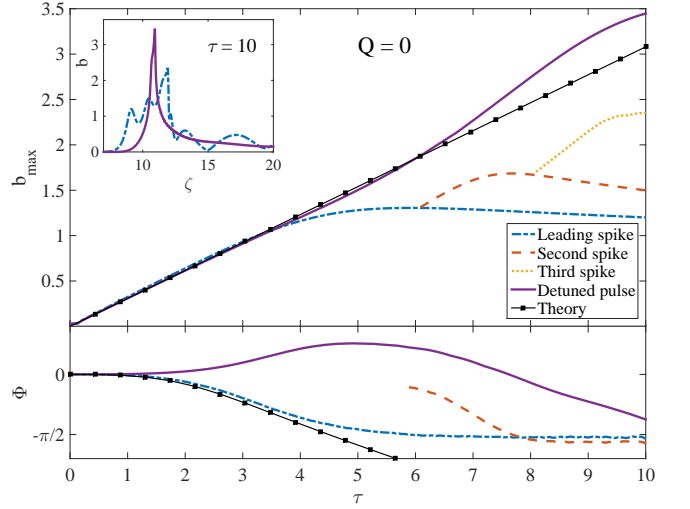}
	\caption{(color online) The maximal amplitude, $b_\text{max}$, 
	(upper panel) and the phase mismatch, $\Phi$, (lower panel)  
	of the amplified pulse  as a function of the amplification time, 
	$\tau$, in the dispersionless $(Q=0)$ regime. 
	$b_\text{max}$ and $\Phi$ of the leading (dashed-dotted), second (dashed) and third (dotted) spikes of the saturated unchirped $(\Delta=0)$ pulse are compared with those of the detuned pulse (solid line) and with the 
	theoretical predictions (solid lines with squares), \Eq{Bmax} for $b_\text{max}$ and \Eq{Phi_theory} for  $\Phi$. 
	The final $\tau=10$ profiles of the singled-spike detuned (solid) 
	and multi-spike undetuned (dashed-doted) are compared in the inset. }
\label{fig1}
\end{figure}

First, let us recall in \Fig{fig1} the nonlinearly saturated (multi-spike)  solution for extremely under-critical plasmas as found by Malkin \etal~\cite{Malkin_PoP_14,Malkin_PRE_14}. 
We solve Eqs.~(\ref{a_}--\ref{b_}) with no detuning, $\Delta=0$, and neglect group velocity dispersion, $Q=0$.
Then, we calculate the total phase mismatch, $\Phi(\zeta,\tau)= \theta-\psi -\varphi$, where, the real amplitudes ($A$, $B$, and $C$) and phases ($\theta$, $\psi$, and $\varphi$) are defined via $a=Ae^{i\theta}$, $b=Be^{i\psi}$, and $f=Fe^{i\varphi}$.
For  a given amplification time, $\tau$, we find the pulse maximum $b_{\rm max}(\tau)= B(\tau,\zeta=\zeta_{\rm max})=\max_{\zeta} [B(\tau,\zeta)]$, where $\zeta_{\rm max}$ is the location of the maximum.
At this location we calculate the phase mismatch, $\Phi_{\rm max}(\tau) = \Phi(\tau,\zeta=\zeta_{\rm max})$.
For the multi-spike solution (as in the inset of \Fig{fig1}), we similarly define the local maxima, $b_{{\rm max},j}$, maxima location, $\zeta_{{\rm max},j}$, and the phase mismatch, $\Phi_{{\rm max},j}$, where $j=1,2$, and $3$, stand for the leading, second, and third spikes, respectively.

In \Fig{fig1} we plot the evolution of the amplitudes of these maxima, $b_{{\rm max},j}$ (upper panel) and the associated phase mismatches, $\Phi_{{\rm max},j}$, (lower panel).
What is important here is that the nonlinear saturation of the amplified pulse maximum $b_{{\rm max},j}$, (upper panel) is time-correlated with its relativistic phase mismatch, $\Phi_{{\rm max},j}$, (lower panel).
It is clearly seen that the amplification of each spike stops when the phase mismatch at the spike's maximum reaches  $\Phi_{{\rm max},j}=-\pi/2$. 
The reason for this correlation can be seen if we rewrite the three complex Eqs.~(\ref{a_})--(\ref{b_}) as six real equations
\begin{eqnarray}
	A_{\zeta}  &=& -BF \cos\Phi, \qquad       \theta_{\zeta} = \frac{BF}{A}\sin\Phi, \label{A}  \\
	F_{\zeta } &=&  AB \cos\Phi, \qquad \;\;  \varphi _{\zeta} = \frac{AB}{F}\sin\Phi -\Delta F,\label{F} \\
	B_{\tau }  &=&  AF \cos\Phi, \qquad  \;\;  \psi_{\tau}  = \frac{AF}{B} \sin\Phi +B^2.  \label{B}   
\end{eqnarray}
It is clear from \Eq{B}  that when $\Phi=-\pi/2$ the three-wave coupling term in the right hand side of amplitude Eqs.~(\ref{A})-(\ref{B}) vanishes, and thus, the amplification of the seed stops.
Next we develop a theory for the phase mismatch evolution of the first spike $\Phi_{{\rm max},1}$ for small $\tau$.

\section{phase mismatch evolution}
\label{sec3}
Since Eqs.~(\ref{a_}--\ref{b_}) for $\Delta=0$ satisfies 
$\left( \left| a \right|^2 + \left| f \right|^2 \right)_{\zeta}=0 $, 
the envelopes can be expressed by 
\begin{eqnarray}
	A &=& \cos(u/2) \\  \label{AA} 
	F &=& -\sin(u/2) \\  \label{FF}  
	B &=& -\frac{u_\zeta}{2\cos\Phi}. \label{BB}
\end{eqnarray}
By substituting Eqs.~(\ref{AA}-\ref{BB}) into Eqs.~(\ref{A}-\ref{B}) we obtain 
that the seed envelope dynamics is described by
\begin{eqnarray}
	u_{\zeta\tau}+u_{\zeta}\Phi_{\tau} \tan\Phi = \sin u \, \cos^2\Phi,   \label{u} 
\end{eqnarray}
while its phase dynamics is
\begin{eqnarray}
	\psi_\tau=\frac{\sin \Phi \cos \Phi \sin u}{u_{\zeta}}+\left(\frac{u_\zeta}{2\cos\Phi}\right)^2. 
	\label{psi_t}
\end{eqnarray}
In the absence of the relativistic effect (the term $B^2$ in \Eq{B}) and detuning ($\Delta=0$)
accompanied by zero phase (\ie real fields) initial conditions, 
$\theta(\zeta,0)=\phi(\zeta,0)=\psi(\zeta,0)= 0$,
Eqs.~(\ref{A})-(\ref{B}) yield $\Phi = \psi = 0$ for all $\zeta$ and $\tau$. 
In such a case, the second term in left hand side of \Eq{u} vanishes and \Eq{u} becomes the sine-Gordon equation, 
\begin{eqnarray}
	u_{\zeta\tau} = \sin u.   \label{sine_gordon} 
\end{eqnarray}
For initial conditions (\ref{b0}), one obtains the $\pi$-pulse solution 
\begin{eqnarray}
	B = \frac{2\tau}{(\xi+1)\cosh(\xi-\xi_M)}. \label{pi-pulse}
\end{eqnarray}
Here, $\xi= 2\sqrt{\tau \zeta}$, and $\xi_M=\ln\left(4\sqrt{2\pi\xi_M}/\epsilon \right)$ is calculated iteratively,  
where for our definitions $\epsilon = 2 \int{b(\zeta,\tau=0) \,  d\zeta }$ is the seed capacity \cite{Malkin_PRL_99,Malkin_PoP_00,Malkin_PoP_14}.
By substituting $\xi=\xi_M$ one finds the maximum of the amplified pulse 
\begin{eqnarray}
	B_{\rm max} = \frac{2\tau}{\xi_M+1}. \label{Bmax}
\end{eqnarray}
When the relativistic effect is small (\ie the weakly nonlinear regime), the phase mismatch between the envelopes is small, 
$\Phi \approx -\psi \ll 1$. 
In this limit, 
$\cos\Phi \approx 1$ and  $\tan\Phi \approx \sin\Phi \approx \Phi$. 
Therefore, we can neglect the second term in the left hand side of \Eq{u} and approximate the seed envelope as the $\pi$-pulse solution (\ref{pi-pulse}).
Then, \Eq{psi_t} becomes
\begin{eqnarray} \label{Phi_t}
	\Phi_\tau = - \Phi\frac{\sin(u)}{u_\zeta} - \left(\frac{u_\zeta}{2}\right)^2
\end{eqnarray}
and \Eq{u} becomes
\begin{eqnarray}\label{u_zt}
	u_{\zeta\tau}+u_{\zeta}\Phi\Phi_\tau = \sin(u).
\end{eqnarray}
Since at the point of the maximum amplified pulse 
${u_{\zeta\tau}} / {u_\zeta} = {B_{\tau}}/{B}=O(\tau^{-1})$, \Eq{u_zt} can be approximated as $\sin(u)/u_\zeta \approx \Phi \Phi_\tau$.
Then \Eq{Phi_t} reads
\begin{eqnarray}\label{Rel:e_8}
	\Phi_\tau = - \frac{1}{3}(\Phi^3)_\tau - \left(\frac{2\tau}{\zeta_M+1}\right)^2,
\end{eqnarray}
that, in turn, is integrated over $\tau$ to get
\begin{eqnarray}\label{Rel:e_9}
	\Phi+\frac{1}{3}\Phi^3 + \mu = 0, 
\end{eqnarray}
where $\mu=\frac{4 \tau^3}{3(\zeta_M+1)^2}$.
The real solution of \Eq{Rel:e_9} is
\begin{eqnarray}\label{Phi_theory}
	\Phi =  \eta^{\frac{1}{3}} -\eta^{-\frac{1}{3}} ,
\end{eqnarray}
where 
$\eta= - \frac{3\mu}{2} +\sqrt{1+ \frac{9\mu^2}{4} }$.
It is notable that for a given amplification time $\tau$, this solution depends on one parameter only, $\epsilon$, which is determined by initial conditions.
For the example presented in \Fig{fig1}, we have
$\epsilon=0.1$ so $\xi_M=5.5$.  
In the upper panel we plot the analytical solution \eq{Bmax} for the maximal amplitude of the leading spike, $B_{\rm max}(\tau)$,  (black line with squares).
In the lower panel of the figure, we compare our theoretical prediction (\ref{Phi_theory}) for the phase mismatch, $\Phi$, of the leading spike (black line with squares) with simulation results (blue dashed-dotted line).
It is notable, that the theory, which does not contain any adjustable parameter, is in a high agreement with the numerical solution.

\section{The effect of detuning} \label{sec4}
We are now in a position to show how frequency detuning compensates the relativistic phase mismatch, enhances the total amplification, and reshapes the amplified pulse.
Consider detuning either by pump chirping or by density gradient.
However, the relevant regime for our problem is the strongly under-critical plasma, $Q\ll1$, because only in this regime, the dynamics is dominated by the relativistic nonlinear saturation and not by dispersion \cite{Malkin_PRE_14}.
In this regime, $\omega_e \ll \omega_b$, and thus, the overall detuning can be mainly achieved 
by pump detuning rather than via density gradient.

\begin{figure}[t]
\includegraphics
	[trim={2.9cm 0cm 3.5cm 0.5cm},clip,width=8.75cm]
	{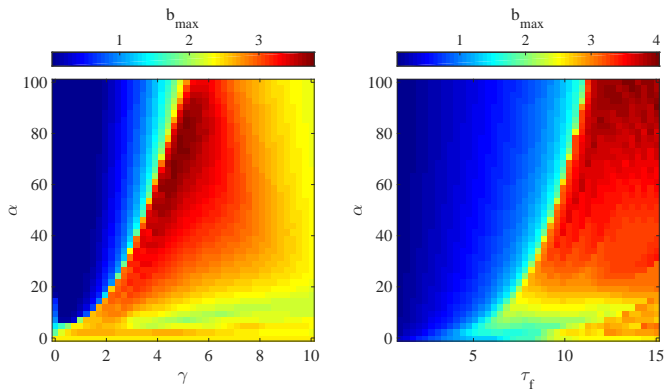}
	\caption{(color online) The maximal amplitude amplification, $b_\text{max} $ as a function 	of the detuning parameters $(\gamma,\alpha)$ for fixed $\tau_f=10$ (left panel) and 	of $(\tau_f,\alpha)$ for fixed 	$\gamma=4$ (right panel) in the dispersionless regime, $Q=0$. }
\label{fig2}
\end{figure}

Following  the theoretical scaling $\Phi \approx -\mu  \sim -\tau^3$ of \Eq{Phi_theory} in the weakly nonlinear stage, $\mu\ll1$ (\ie $\tau<3$ for the example shown in \Fig{fig1}), we consider detuning of the form
\begin{equation}
	\Delta(\tau)= \alpha \left(\frac{\tau}{\tau_f}\right)^\gamma. \label{detuning}
\end{equation}
Here, $\tau_f$ is the total amplification time (proportional to the plasma length) and the values of the parameters $\alpha$ and $\gamma$ are to be optimized.
Optimization over $\gamma$ is also needed, because for larger amplification  times, $\tau_f>4$, different values of $\gamma$ may yield a better results.
For example, in \Fig{fig1} we illustrate the effect of chirping by considering detuning  with parameters $\alpha=55$ and $\gamma=4$. 
This value of $\gamma$ was chosen because, in this case, where $\tau_f=10$, it results in a larger pulse intensity and better pulse profile than $\gamma=3$ (see also \Fig{fig2}). 
The spikes maxima of the aforementioned unchirped solution (dashed and dashed-dotted lines) are compared in \Fig{fig1}, with the maximum of the chirped solution (solid line).
The final $(\tau_f=10)$ amplified (chirped and unchirped)  profiles, $|b(\zeta,\tau=\tau_f)|$ are compared in the inset.

We note that the chirped solution is better than the unchirped solution in two aspects.
First, the maximal amplitude of the amplified pulse increases by about $50\%$ so the intensity is enhanced by more than a factor of two.
Moreover, the pump detuning reshapes the multi-spike unchirped  solution into a single spike pulse, resulting in a higher intensity in the leading spike and less precursory power ahead of it.
For the example shown in \Fig{fig1}, the leading spike maximum increases  by a factor of $2.6$ so the maximal intensity of the leading spike increases by more than a factor of $6$ without additional precursory spikes. 

This reshaping effect might be essential for experiments in which contrast ratio plays an important role.
As can be seen in \Fig{fig1} (upper panel), the leading spike maximum of the detuned solution agrees with the analytical expression of \Eq{Bmax} for the $\pi$-pulse solution also for $\tau>4$, when the unchirped solution is already saturated.
This agreement, accompanied by the pulse reshaping, support the understanding that the pump detuning maintains the amplification of the $\pi$-pulse solution beyond the (unchirped) nonlinear saturation limit.
It is also notable that a negative chirp $(\alpha<0)$ has the opposite effect, \ie increasing the phase mismatch, $\Phi$, reducing the pulse amplification (not shown in the figures). 

Obviously, one would want to optimize the detuning with respect to a certain objective function;
some examples are the maximum output intensity, the amplified pulse profile, and the energy transfer efficiency.
To this end, one should define a detuning profile and optimize (\eg via a genetic algorithm) its parameters with respect to an objective function.
For simplicity, we choose here to look at the amplified pulse maximum, $b_\text{max}$, and study its dependency on the detuning parameters $\alpha$ and $\gamma$ of the detuning profile (\ref{detuning}) and the total amplification time, $\tau_f$.
Of course, different objective functions or detuning profiles would be suitable for specific experimental goals and restrictions.
However, these further calculations are outside the scope of this paper.

In \Fig{fig2}, we present the maximum amplitude, $b_{\rm max}$ in the 
$(\gamma,\alpha)$ parameter space (left panel) for constant $\tau_f=10$, and in the $(\alpha,\tau_f)$ plane for constant $\gamma=4$ (right panel).
The sharp transitions between very low ($b_\text{max}<1$) and large ($b_\text{max}>3$) amplifications in both panels indicate that for detuning that is too abrupt (too small $\gamma$ or too small $\tau_f$ for a given $\alpha$), the Raman resonance condition is lost at early times, resulting in an insignificant  amplification.
This effect is important to mitigate the premature noise amplification as	suggested in Ref. \cite{Malkin_PRL_00_a}.
In addition to the slowness of the detuning, a minimum detuning amplitude, $\alpha$, is needed for significant pulse amplification.
For the example of \Fig{fig2} one finds the condition $\alpha\ge15$.
Note that pulse reshaping is not included in the figure. 
Actually, although the example of \Fig{fig1} has only $b_{\rm max}\approx 3.5$ while the maximum $b_{\rm max}$ in \Fig{fig2} is about $4$, it has a mono-spike  pulse with 
less precursory power.

\section{Detuning in dispersive medium}\label{sec5}

\begin{figure}[b]
	\includegraphics
	[width=9cm]
	{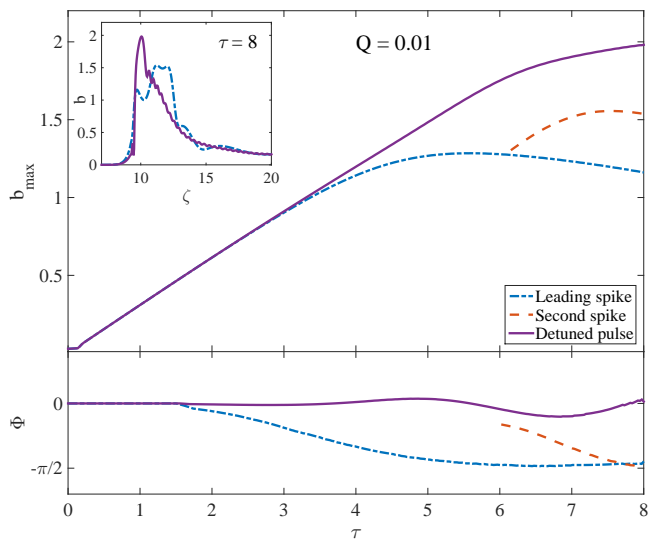}
	\caption{(color online) The maximal amplitude, $b_\text{max}$, 
	(upper panel) and the phase mismatch, $\Phi$, (lower panel)  
	of the amplified pulse  as a function of the amplification time, 
	$\tau$ in the finite dispersion $(Q=0.01)$ regime. 
	The saturation of the leading (dashed-dotted) and second (dashed) 
	spikes of the unchirped $(\Delta=0)$ pulse 
	are compared with the maximal amplitude of the chirped pulse (solid line). 
	The final $\tau=8$ profiles of the singled-spike chirped (solid) 
	and multi-spike unchirped (dashed-doted) are compared in the inset.}
	\label{fig3}
\end{figure}

\begin{figure}[t]
\includegraphics
	[trim={2.9cm 0cm 3.5cm 0.5cm},clip,width=8.75cm]
	{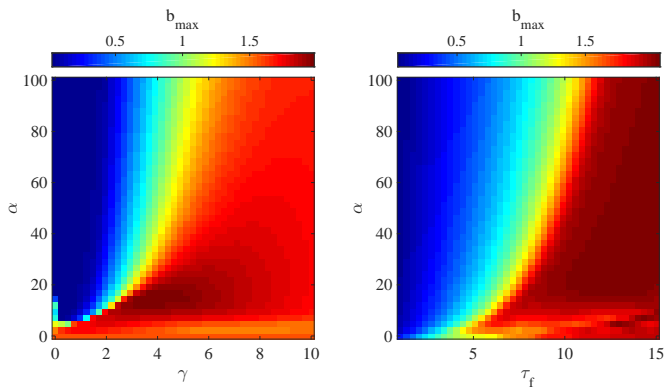}
	\caption{(color online) The maximal amplitude amplification, $b_\text{max}$, as a function 
	of the detuning parameters $(\gamma,\alpha)$ for fixed $\tau_f=8$ (left panel) and of
	$(\tau_f,\alpha)$ for fixed $\gamma=4$ (right panel) 
	in the finite dispersion regime, $Q=0.01$.}
\label{fig4}
\end{figure}

Next we study the case of non-negligible, but small, dispersion \ie $Q\ll1$.
We solve Eqs.~(\ref{a_}--\ref{b_}) for the same initial conditions as in the dispersionless ($Q=0$) case, but now with $Q=0.01$.  
Similarly to \Fig{fig1}, we compare in \Fig{fig3}, the unchirped, multi-spike solution (dashed and dashed-doted lines) with the chirped solution (solid line).
The detuning parameters here are $\delta=4$, $\alpha=15$, and $\tau_f=8$.
In the detuned solution the pulse amplification is significantly higher than in the undetuned solution and the maximum intensity is achieved at the pulse front without precursory spikes. 
However, the effect here is modest compared to the dispersionless case due to the dispersive broadening of the pulse.
As before, we illustrate, in  \Fig{fig4}, an optimization roadmap by plotting $b_{\rm max}$ as a function of the detuning parameters $\alpha$ and $\gamma$ and the amplification time $\tau_f$.
A comparison with \Fig{fig2} reveals that for finite $Q$, smaller detuning amplitudes, $\alpha$, are need  for optimal amplification.
Note that these maps show only part of the whole picture, because neither pulse reshaping nor efficiency consideration were taken into account.
For example,  although for $\tau_f>10$ (in the right panel) one can find a solution with higher $b_\text{max}$ than shown in \Fig{fig3} where $\tau_f=8$, it will be a multi-spike pulse (not shown) so the amplified pulse profile is not as good as in \Fig{fig3}.
We also found that the pump detuning is more effective for small values of $Q$.
This is because larger dispersion broadens the amplified pulse, so $b_{\rm max}$ saturates due to dispersion rather than by relativistic nonlinearity.
Therefore, this saturation cannot be overcome by pump detuning.

One might imagine that we may combine seed chirping \cite{Toroker_PRL_12} with pump detuning in order to reduce the amplification time and therefore to improve the efficiency of the detuned BRA.
However, these two methods operate in different regimes.
Seed chirping is useful for high densities ($\omega_e>0.25 \omega_b$), where the group velocity dispersion of the seed pulse becomes a dominant effect (see Fig.~1 in Ref.~\cite{Toroker_PRL_12}).
On the other hand, pump detuning is effective for small densities ($\omega_e<0.1 \omega_b$), where the group velocity dispersion does not shadow the relativistic nonlinearity (see Fig.~3 in Ref.~\cite{Malkin_PRE_14}).

Finally, let us consider what could be the parameters for a future experiment.
For example, for seed wave length $\lambda_b \approx \lambda_b =0.1 \, \mu {\rm m}$,  
and $Q= 0.01$ (corresponding to $\omega_e/\omega_b = 0.02$), the plasma density is $n_e = 4.4 \times 10^{19} \,  {\rm cm}^{-3}$.
In this case, for
pump intensity $I_0 = I_{\rm br}/2$, where $I_{\rm br}$ is the wave breaking threshold (see \Eq{I_br}), we have
$I_0 = 6.7 \times 10^{13}  \, {\rm W/cm}^2$.
For circular polarized field (\ie $p=2$), the dimensionless pump amplitude is given by
$a_0 =  6 \times 10^{-10} \, \lambda_a [\mu \rm m] \sqrt{I[{\rm W/cm}^2]}  =  5 \times 10^{-4}$.
Therefore, for the example shown in \Fig{fig3}, the pump duration is $0.85 \, {\rm ns}$, which corresponds to plasma length of $12.75 \, {\rm cm}$, while the total detuning  of the pump frequency is $7.5\%$.
Without pump detuning, the maximal amplified intensity is achieved in the second spike and equals to $3.1 \times 10^{18} \,  {\rm W/cm}^2$ while the first spike intensity is $1.8 \times 10^{18}  \, {\rm W/cm}^2$. 
The detuning in this example improves the maximal intensity of the first (and only) spike up to $5.2 \times 10^{18}  \, {\rm W/cm}^2$ with output duration (full width half maximum intensity) of $12 \, {\rm fs}$.
Notably, although proper detuning significantly increases the output intensity, the fluence and efficiency remain approximately the same.
For no detuning ($\Delta=0$), the fluence, that is defined in \Eq{w_b}, is $w_b=8.5 \times 10^4 {\rm J/cm}^2$, where the integration was taken between $\zeta=5$ and $\zeta=15$, and the efficiency (see \Eq{efficiency}) is $\eta_w=0.75$. 
Employing detuning in this example changes these values by less then $3\%$.
For the same pump intensity, but with linear polarization ($p=1$), shorter plasma but larger detuning are required to get the same amplification.

\section{secondary Raman backscattering}\label{sec6}
The amplification also tends to saturate due to the secondary Raman backscattering (SRBS) of the amplified pulse, $b$, into downshifted, counter propagating noise of frequency $\omega_b-\omega_e$.
It was shown in Ref. \cite{Malkin_PRL_99} that, for the $\pi-$pulse solution, this effect is minor  (about 5 exponentiations only) and does not deplete much of the amplified pulse.
Nevertheless, in the relativistic regime the amplified pulse is longer so there is more time for SRBS.
Also, the group velocity dispersion broadens the pulse, so it becomes more susceptible to SRBS as $Q$ increases.  
We define the SRBS total noise amplification, $e^\Gamma$, where (see \Eq{Gamma})
\begin{equation} \label{Gamma1}
	\Gamma= \sqrt{\sigma} \int |b| d\zeta,
\end{equation}
and for strongly under-critical plasma, $\sigma\approx 2$.
For the dispersionless profiles in the inset of \Fig{fig1}, one finds $\Gamma = 7.8$ for the unchirped (dashed line), which reduces to $\Gamma = 6$ for the chirped solution (solid line). 
Similarly, but less significantly, for the examples of \Fig{fig3} in the dispersive regime,
one finds $\Gamma = 7.5$ for the unchirped solution (dashed line in the inset of \Fig{fig3}) and $\Gamma = 7$ for the chirped solution (solid line).
In both cases, the integration range in \Eq{Gamma1} was $5<\zeta<15$.

Next we estimate the initial (thermal) noise intensity in order to  determine the significance of this effect in experiment.
To this end, we consider electron temperature $T_e$, with associated electric field $E_{\rm noise}=T_e/e\lambda_D$, where $-e$ is the electron charge and $\lambda_D [{\rm cm}] = 743\sqrt{T_e[eV]/n_e[{\rm cm}^{-3}]}$ is the Debye length.
Hence, the electric field can be written as 
$E_\text{noise}[{\rm V/cm}]=10^{-3} \sqrt{T_e[eV] n_e[{\rm cm}^{-3}]}$
and the corresponding noise intensity is
\begin{equation}
	I_\text{noise}[{\rm W cm}^{-2}]=10^{-9} \, T_e[{\rm eV}] \, n_e[{\rm cm}^{-3}].
\end{equation}
Here we assume $\delta$-correlated noise with flat spectrum and a cut-off at $\lambda_D/c$.
The Raman instability width can be estimated as the SRBS increment, $\tilde{b}_{m} V_3$, where $V_3$ is the three wave coupling that is defined in \Eq{V3} and $\tilde{b}_{m}$ is a typical seed amplitude (see \Eq{b1}). 
Therefore, the effective noise intensity reads
\begin{equation}
	I_\text{eff} =   I_\text{noise} \, \frac{\tilde{b}_m V_3\lambda_D}{c}.
\end{equation}
Finally, for $\lambda_b=0.1\mu {\rm m}$, $n_e=10^{19} {\rm cm}^{-3}$, $T_e=100 {\rm eV}$, and typical amplified pulse amplitude $b_m=1.5$ (\ie $\tilde{b}_{m}=0.1$), one finds
$V_3=1.9 \times 10^{15}\, {\rm sec}^{-1}$ and 
$\lambda_D=1.1 \times 10^{-6}\, {\rm cm}$.
For this example, the total background noise can be estimated as
$I_\text{noise} = 4.4\times 10^{12} \,   {\rm W/ cm}^{2}$ 
while the effective noise intensity available to SRBS would be
$I_\text{eff}      = 3.3  \times 10^{10} \, {\rm W/ cm}^{2}$.
This means that $18$ exponentiations separate the effective noise intensity and the amplified pulse intensity that was estimated in the previous section to be of the order of $3 \times 10^{18} \, {\rm W/ cm}^{2}$. 
For the aforementioned example the number of SRBS intensity exponentiations is only $2\Gamma\approx14$, so the SRBS is predicted to be harmless in this case.

\section{Summary} \label{sec7}
In summary, we found that pump detuning can mitigate the relativistic nonlinear saturation of the leading spike for strongly under-critical plasmas. 
This occurs when detuning compensates the relativistic phase mismatch  that causes the saturation of the leading spike. 
The benefits of this compensation are twofold.
First, the amplification of the maximal intensity can be enhanced by as much as a factor of two compared to the achievable amplification without pump detuning.
Second, the amplified pulse is reshaped into a single-spike pulse with significantly less precursory power ahead of it.
Also, the reshaping of the leading spike reduces the effect of the secondary Raman backscattering of the amplified pulse. 
The precise pulse reshaping and maximum intensity were shown to depend upon the precise detuning parameters.
It is worth noting that the technique proposed here for overcoming the nonlinear saturation is not limited to BRA but is, in fact, a universal solution of the three wave interaction problem in various physical systems.

\acknowledgments{
This work was supported by 
NNSA Grant No.~DE-NA0002948, 
AFOSR Grant No.~FA9550-15-1-0391, 
DOE Contract No.~DE-AC02-09CH11466, 
DTRA Grant No.~HDTRA1-11-1-0037, 
RFBR Grant No.~15-32-20641, 
and by the Dynasty Foundation.
}


\appendix*
\section{Derivation of the model}

This appendix present a derivation of Eqs.~(\ref{a_})--(\ref{b_}).
We begin with the usual three wave interaction equations including nonlinearity and detuning within the fluid model \cite{Malkin_PRL_00_a,Malkin_PRL_07,Malkin_PRE_14}
\begin{eqnarray}
	\tilde{a}_t +c_a \tilde{a}_z &=& V_3 \tilde{f} \tilde{b}  \label{a1_t}  \\
	\tilde{b}_t - c_b \tilde{b}_z &=&-V_3 \tilde{a} \tilde{f}^* + i R | \tilde{b} |^2 \tilde{b} - i \kappa   \tilde{b}_{tt}\\
	\tilde{f}_t  + i \delta\omega \tilde{f} &=& - V_3 \tilde{a} \tilde{b}^{*}.  \label{f1_t}
\end{eqnarray}
Here, $\tilde{a}$ and $\tilde{b}$ are the vector-potential envelopes of the pump pulse and the counterpropagating shorter pumped pulse, respectively, measured in units of
$m_e c^2/e \approx 5\times10^5 V$; 
$\tilde{f}$ is the rescaled envelope of the Langmuir wave electrostatic field
in units of $(m_e c/e) \sqrt{p\, \omega_e \omega_a} $.  
Here, $\omega_{a,b}$ are the frequencies of the pump and the seed pulses.
$c$ is the vacuum speed of light; 
$\omega_e=4\pi e^2 n_e /m_e$ is the plasma frequency;
$-e, m_e$, and $n_e$ are the electron charge, mass, and density, respectively;
Subscripts $t$ and $z$ signify time and space derivatives and
\begin{equation}
	c_{a,b}=c\sqrt{1-(\omega_e/\omega_{a,b})^2}
\end{equation}
are the group velocities of the pump and the seed.
The parameter $p$ determines the polarizations of the pulses, where
\begin{eqnarray}
	p &=& 1            \qquad  \text{ Linear polarization} \\
	p &=& 2            \qquad \text{ Circular polarization}.
\end{eqnarray}
The 3-wave coupling constant, $V_3$ (real for appropriately defined wave envelopes),
can be written as \cite{Kruer,Clark_PoP_02}
\begin{equation} \label{V3}
	V_3=k_f c \sqrt{\frac{p\, \omega_e}{16\omega_b}}, 
\end{equation}
where $k_f$ is the wave number of the resonant Langmuir wave, \ie
\begin{eqnarray}
	k_f       &=& k_a + k_b, \\
	k_{a,b} &=& \frac{\omega_{a,b}}{c}  \sqrt{1 - \frac{\omega_e^2}{\omega_{a,b}^2}}.
\end{eqnarray}
The corresponding frequency resonance condition is
\begin{equation}
	\omega_b+\omega_f =\omega_a, \label{resonance}
\end{equation}
where $\omega_f \approx  \omega_e$ is the Langmuir wave frequency in a cold plasma. 
For under-critical plasma, $\omega_e\ll\omega_a$, we can approximate $\omega_a \approx\omega_b$ and $k_f\approx 2k_a$, so $V_3\approx  \sqrt{p\, \omega_e\omega_a}\, /2$.
Then, the nonlinear frequency shift coefficient due to the relativistic electron nonlinearity, $R$, reads \cite{Litvak, Max, Sun}
\begin{equation}
	R=\frac{p\,\omega_e^2\omega_a}{8\omega_b^2}\approx \frac{p\,\omega_e^2}{8\omega_a}.
\end{equation}
The group velocity dispersion coefficient is
\begin{equation}
	\kappa = \frac{1}{2 c_b} \frac{d c_b}{d\omega} 
	=\frac{\omega_e^2 c^2}{\omega_b^3 c_b^2}
\end{equation}
and the total detuning from the perfect Raman frequency resonance condition 
(\ref{resonance}) is given by
\begin{equation}
	\delta\omega=\omega_f+\omega_b-\omega_a.
\end{equation}
Note that, in this normalization, the average square of the electron quiver velocity in the pump laser field, measured in units of $c^2$, is $|\tilde{a}|^2$, is such that $v_{ea}^2=c^2|\tilde{a}|^2$.
Also, the average square of the electron quiver velocity in the seed laser field and in the Langmuir wave field are given by $v_{eb}^2 = c^2 |\tilde{b}|^2 \frac{\omega_a}{\omega_b}$ and $	v_{ef}^2  = c^2 |\tilde{f}|^2 \,p\,\frac{\omega_a}{\omega_f}$, respectively.   
This hydrodynamic model is applicable for pump pulse
intensities $I_a$ smaller than the Langmuir wave breaking threshold $I_\text{br}$,
\begin{equation} \label{I_a}
	I_a = \frac{\pi \, m_e^2 c^5}{e^2\lambda_a^2}  \frac{p}{2} \, |\tilde{a}|^2
	\approx \frac{5.48 \times10^{18}\,|\tilde{a}|^2}{\left(\lambda_a[\mu{\rm m}]\right)^{2}} \frac{p}{2} \;  \frac{\rm W}{{\rm cm}^2}.
\end{equation}
If we assume nearly complete pump depletion, then $I_\text{br}$ can be written as \cite{Toroker_PoP_14}
\begin{equation} \label{I_br}
	I_\text{br}=\frac{m_e n_e c^3 \omega_e}{16 \omega_a}.
\end{equation}
Therefore, the wave breaking vector potential for both linear ($p=1$) and circular ($p=2$) polarizations is  
\begin{equation} \label{a_br}
	a_\text{br}=\sqrt{\frac{2 e^2 \lambda_a^2 I_\text{br}}{p\,\pi m_e^2 c^5}}
	= \frac{1}{\sqrt{p}} \left(\frac{\omega_e}{2\omega_a} \right)^{\frac{3}{2} }  
	\approx \frac{ Q^{3/2}}{\sqrt{p}},
\end{equation}
where, 
 $\lambda_a=2\pi/k_a$ is the pump wavelength, and $Q\approx q/2=\omega_e/2\omega_b$ (see below).
Next, we adopt a universal model \cite{Malkin_PRE_14} by transforming to the dimensionless variables
\begin{eqnarray}
	\tau  &=& \left(\sigma R V_3^2 a_0^4\right)^{\frac{1}{3}} \frac{L-z}{c_b}  \label{tau}\\
	\zeta &=& \left(\frac{ V_3^4 a_0^2}{\sigma R}\right)^{\frac{1}{3}} \left(t- \frac{L-z}{c_b}\right)  \label{zeta}
\end{eqnarray}
where $\sigma=1+\frac{c_a}{c_b}\approx 2$.
Here, $\tau$ measures the elapsed amplification time (or the distance traversed by the original seed front); $\zeta$ measures the distance (or delay time) from the original seed front; $L$ is the plasma length and $a_0$ is the input pump amplitude; the seed is injected into the plasma at $z=L$, $t=0$ and meets immediately the pump front injected into the plasma at 
$z=0$, $t=-L/c_a$. 
Then, we define new wave amplitudes, $a,b$, and $f$, via
\begin{eqnarray}
\tilde{a} &=& a_0 \,a,  \label{a1} \\
\tilde{f}  &=& -a_0  \sqrt{\sigma}  \,f,  \label{f1} \\
\tilde{b} &=& \left(\frac{V_3 a_0^2 \sqrt{\sigma}}{R}\right)^{\frac{1}{3} }  \, b. \label{b1}
\end{eqnarray}
By neglecting the ``slow" time derivative of the pump amplitude compared to the ``fast" time derivative of the pump amplitude, Eqs.~(\ref{a1_t}--\ref{f1_t}) read
\begin{eqnarray}
a_{\zeta} &=& -b f,       \label{a_tilde} \\
f_{\zeta } &=& a b^{*} - i\Delta f,  \label{f_tilde} \\
b_{\tau } &=& a f^{*} - iQ \, b_{\zeta\zeta} + i |b|^2 \, b.  \label{b_tilde} 
\end{eqnarray}
Here, 
\begin{equation}
	\Delta = \left(\frac{\sigma R }{V_3^4 a_0^2}\right)^{\frac{1}{3}}\delta\omega
\end{equation}
is the rescaled detuning and  the parameter
\begin{equation} \label{Q}
	Q=\frac{(k_a+k_b)^2c^2\omega_b}{4\omega_e\omega_a(c_a+c_b)}\frac{dc_b}{d\omega}
\end{equation}
characterizes the group velocity dispersion of amplified pulse and depends only on the ratio of the plasma to laser frequency $q = \omega_e / \omega_b$ .
In strongly under-critical plasmas,
where $q \ll 1$, one finds $Q = q/2$ and $\sigma=2$; 
in nearly critical plasmas, where $q \to 1$, one has $Q = 0.5/\sqrt{ 1 - q^2} \gg 1$.
With Eqs.~(\ref{a_tilde})--(\ref{b_tilde}) we arrive at Eqs.~(\ref{a_})--(\ref{b_}) that comprise the physical model of this paper.
In strongly under-critical plasmas (\ie $Q\approx \omega_e/2 \omega_b\ll1$ and $\sigma\approx 2$), which is of major interest here, the amplified pulse intensity,  $I_b$, can be expressed in these variables as
\begin{eqnarray}
	I_b =  \frac{I_0 |b|^2}{Q} \left(\frac{4}{p\, a_0^2}\right)^{\frac{1}{3}} 
	= \frac{G \, \omega_e |b|^2}{4 \, \lambda_b}\left(\frac{2 I_0^2}{p^2 I_\text{br}^2}\right)^{\frac{1}{3} }, 
\end{eqnarray}
where $G=m_e^2c^4/e^2=0.3 \, {\rm J/cm}$ as in Refs. \cite{Malkin_PoP_14,Malkin_PRE_14}.
In this limit, the final amplified pulse fluence can be calculated at time $\tau=\tau_f$ via
\begin{eqnarray} \label{w_b}
	w_b = \int I_b dt 
	= \frac{I_0}{Q \,\omega_a } \left(\frac{16}{p^2  a_0^4 }\right)^{\frac{1}{3}} \int |b|^2 d \zeta. \end{eqnarray}
For constant pump, $I_a=I_0$, the total pump fluence that was invested in the system, 
$w_a = I_0 \Delta t_a$, where $\Delta t_a = 2L/c$ is the pump duration, can be written as
\begin{eqnarray} 	\label{w_a}
	w_a = \frac{2 \tau_f I_0 }{Q\, \omega_a}\left(\frac{2}{p^2 a_0^4}\right)^{\frac{1}{3}} 	= \frac{G \tau_f}	{\lambda_a}\left(\frac{I_0}{4I_{\rm br}}\right)^{\frac{1}{3}}.
\end{eqnarray}
Then, the efficiency can be defined as \cite{Edwards_PoP_15}
\begin{equation} \label{efficiency}
	\eta_w = \frac{w_b}{w_a} = \frac{1}{\tau_f} \int |b|^2 d \zeta,
\end{equation}
where we assumed that the initial energy in the seed is small compared to the final energy in the amplified pulse.

Finally, we note that  the number of exponentiations of the SBRS (see \Sec{sec6})
can be written in these rescaled variables as
\begin{equation} \label {Gamma}
	\Gamma=V_3 \int |\tilde{b}| dt = \sqrt{\sigma} \int |b| d\zeta.
\end{equation}


\end{document}